\documentclass[10pt,pre,superscriptaddress,aps]{revtex4-1}
\usepackage{amsmath,amssymb}
\usepackage{graphics,graphicx}
\usepackage{color}

\begin{document}

\title{Ballistic annihilation in one dimension : A critical review}

\author{Soham Biswas}%

\affiliation{Departamento de F\'isica, Universidad de Guadalajara, Guadalajara, Jalisco, Mexico }

 \email{soham.biswas@academicos.udg.mx}

\author{Francois Leyvraz}

\affiliation{Instituto de Ciencias F\'isicas, Universidad Nacional Aut\'onoma de M\'exico, Mexico }

 \email{f_leyvraz2001@hotmail.com}
 
\affiliation{ Centro de Ciencias F\'isicas, Cuernavaca, Morelos, Mexico}

\begin{abstract}
In this article we review the problem of reaction annihilation $A+A  \rightarrow \emptyset$ on a real lattice in one dimension, 
where $A$ particles move ballistically in one direction with a discrete set of possible velocities.  
We first discuss the case of pure ballistic annihilation, that is a model in which each particle moves simultaneously at constant speed. 
We then review ballistic annihilation with superimposed diffusion in one dimension. This model consists of diffusing particles 
each of which diffuses with a fixed bias, which can be either positive or negative with probability $1/2$, and annihilate upon contact.  
When the initial concentration of left and right moving particles is same, the concentration
$c(t)$ decays as $t^{-1/2}$ with time, for pure ballistic annihilation.
However when the diffusion is superimposed decay is faster and the concentration $c(t) \sim t^{-3/4}$. We also discuss 
the nearest-neighbor distance distribution as well as crossover behavior.  
\end{abstract}

\maketitle

\section{Introduction}
\label{sec:intro}
The dependence of chemical kinetics on the dynamics of the reacting particles has attracted  considerable attention
\cite{review, book}. In all these works, one of the most important objective is to determine how the laws of motion 
influence the space-time evolution of the species concentrations. For irreversible diffusion-controlled reactions, when reactants diffuse and react on
contact, the reaction creates spatial correlations that invalidate the mean-field approximation  and hence the applicability of rate equations. 
For instance, consider the case of one-species annihilation (or aggregation) where one species reacts with itself via the bi-molecular reaction,
\begin{equation}
A+A \mathop{\longrightarrow}_K C ,
\label{eeq:1}
\end{equation}
where $C$ is an inert species.
The rate equation reads : 
\begin{equation}
\dot c_A=-Kc_A^2,
\label{eq:1new}
\end{equation}
 According to the rate equation the $A$  concentration $c_A(t)$ decays as $1/t$ at long times. However for the same process
where $A$ walkers diffuse randomly in one dimension,
the concentration decay is given by $(Dt)^{-1/2}$ \cite{spouge}. Also the amplitude in Eq.(\ref{eq:1new}) depends 
on the initial concentration, whereas the the amplitude of the correct 
$1/\sqrt{t}$ decay only depends on the diffusion constant.

Beyond the immediate interest of such models for chemical reactions, the dynamical behaviour of  particle  annihilation 
($A+A \rightarrow \emptyset$) and particle aggregation ($A+A \rightarrow A$) 
have a direct one-to-one correspondence with the kinetic Ising and 
Potts model in one dimension. In that case, the zero-temperature Glauber dynamics can be mapped 
to a random walk problem for both the $q$ state nearest neighbour Potts model and Ising model ($q=2$). 
In both cases domain walls perform random walks, and whenever the walkers meet, they either 
coagulate ($A + A \rightarrow A$) or annihilate ($A + A  \rightarrow \emptyset$). Two domain walls
will annihilate each other by contact, if the two spins that surround that domain wall are in same state.
Specifically, for the Ising model ($q=2$ state Potts model) the domain walls always annihilate. For the $q$-state Potts model with random initial conditions, 
they will annihilate with a probability $\dfrac{1}{q-1}$ and will coagulate with probability $\dfrac{q-2}{q-1}$.
In either case the domain walls perform pure random walks.
The exponent for the concentration decay is identical to that of the domain growth exponent \cite{Francois} 
or ordering exponent \cite{bnnni}. However for several binary opinion dynamic models 
(which can  be  directly  mapped  to the Ising spin system) in one dimension, the corresponding walker pictures 
are much more complicated than a simple random walk. Sometimes the motion of the $A$ walkers can also 
have ballistic components \cite{opn}. Hence understanding the ballistic annihilation with and without superimposed 
diffusion could also be useful for understanding these more complex dynamics. 

Here in this article we review the problem of ballistic annihilation in one dimension. The main aim is to 
determine the decay law for the concentration of the particles for this single species 
reaction-annihilation problem. First, in Section II we discuss the ballistic annihilation ($A+A \rightarrow \emptyset$) 
where the $A$ particles move in a purely ballistic way on the real line with velocities which can take only two values, 
$v$ and $-v$. We assume the particles to have always 
one of a discrete set of velocities. This model can also be simulated on a lattice, if we assume that all particles always move in 
one direction and if their positions are updated {\em synchronously}. We shall thus refer to this model as the synchronous update model. 
In section III, we talk about the ballistic annihilation with superimposed diffusion in one dimension. This model can be obtained, for instance,
by modifying the update rule of the previous model to being asynchronous. The conclusion and discussions are 
stated in the final section. 

\section{Synchronous Ballistic annihilation}
\label{sec:ballsync}
\subsection{Particles with only two velocities $+v$ and $-v$}
\label{subsec:2v}
The purely ballistic model, that is the Ballistic annihilation under the synchronous update rule was first 
discussed by Elskens and Frisch \cite{balsyn}. 
In this model, particles are initially distributed randomly on a one dimensional lattice and move freely with 
independent initial velocities. In this single species annihilation problem 
whenever two particle meet they annihilate ($A+A \rightarrow \emptyset$). The 
velocity distribution function, that is, the probability that  the velocity is $v$ or $-v$,can be written as 
\begin{equation}
D(w)=q\delta(w +v) + p\delta(w - v),\qquad p+q=1
\label{veldist}
\end{equation}
The main interest is to determine the decay law for the concentration decay of the particles.

We will now look at the problem in a little more detail. Initially let the position and velocity of the $k$th particle be $x_k$ and $v_k$. The 
particle positions are initially random
and the velocities are independent. What we want to calculate is the average \textit{survival probability} $S(t)$ of the 
particles at time $t$, as the concentration decay is identical to the decay of survival probability with time. 
 
 \begin{figure}[ht]
 \includegraphics[width=5cm,angle=0]{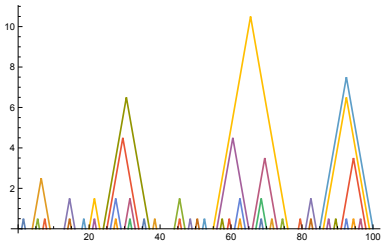} 
  \caption{This shows a schematic picture for the space--time evolution of the system of particles going through the 
  process of pure ballistic annihilation, where the 
  distribution of the velocities are given by equation (\ref{veldist}),
  that is,  where the particles can have only two velocities $+v$ or $-v$. 
  Vertical direction is the direction of time. }
  \label{schematic_syn}
\end{figure}
 
In this case two particles which are moving in the same direction move with the same velocity and thus cannot annihilate
(Figure. \ref{schematic_syn}a). Since, because of the annihilation process, particles cannot cross each other's trajectories , 
it follows that the entire distribution of velocities of surviving particles to the right 
of a given surviving particle (let us denote the index by $k$) is independent of the whole distribution to its left. 

The time at which any given particle, say a right moving particle, will annihilate is readily determined
as follows, if the initial conditions are given: let us denote the given particle by the index 0 and place it
at the origin.  We further assign 
the index $k$ to the $k$'th particle to the right which is located at $x(k)$, and compute
\begin{equation}
S(k)=\sum_{l=0}^k \rm{sign} (v_k)
\label{eq:2.1}
\end{equation}
Denote by $k_0$ the smallest value of $k$ such that $S(k_0)=0$. Note that $k_0$
is necessarily even. We then find that
the annihilation time for the particle 0 is given by
\begin{equation}
T_0=x(k_0)/(2c)
\end{equation}
 The probability density $p(t)dt$ for $T_0$ is thus given by
 \begin{widetext}
 \begin{equation}
p(t)\,dt=\sum_{k=1}^\infty\mbox{\rm Prob}\left[
S(k)=0\mbox{\rm\ for the first time}
\right]\cdot\mbox{\rm Prob}\left[
2vt\leq\sum_{l=1}^k \left(
x_{l+1}-x_l
\right)\leq 2c(t+dt)
\right]
\label{eq:2.2}
\end{equation}
\end{widetext}
This sum is readily evaluated as follows: the first factor in the product is a well-known problem in probability, 
it is the so-called gambler's ruin problem, whereas the second is simply the expression for the distribution of the
distance of the $k$'th neighbour to the origin, given the nearest neighbor distance distribution.  
By the law of large numbers, if this distribution has a finite mean $\langle l\rangle$ and variance $\sigma^2$, the second factor, 
for large $k$, is a Gaussian of average $k\langle l\rangle$ and variance $k\sigma^2$. 

The behaviour of the first factor is somewhat non-trivial but well-known, see for instance \cite{feller}. When $p=q=1/2$, the 
asymptotic  behaviour of this quantity goes a a power law $k^{-3/2}$. It corresponds to the probability that an unbiased
random walk remains to the right of the origin for $k$ steps. On the other hand, if $p>q$, the random walk is biased and 
with probability one eventually remains forever on the right side of the origin, in other words, right-going particles have a 
finite probability of surviving indefinitely.

For $p=1/2$ and an initial exponential nearest neighbor distance distribution, the expression [\ref{eq:2.2}]
can be evaluated without any approximation \cite{balsyn}. The survival probability is given by
\begin{equation}
S(t) = \int_t^\infty p(t^\prime)\,dt^\prime = e^{-2\rho vt}[I_{0}(2\rho vt)+I_{1}(2\rho vt)]
\end{equation}
where $I_{0}$ and $I_{1}$ are the modified Bessel's functions and $\rho=\langle l\rangle^{-1}$. 
Asymptotically for $t\rightarrow \infty$, $S(t)$ decays in a power law fashion:
\begin{equation}
  S(t)= \dfrac{1}{\sqrt{\pi\rho vt}}\left[
  1+O(t^{-1})
  \right]
  \label{survial}
\end{equation}
For $p>q$, on the other hand, the expressions for the probability of a random walk remaining for $k$
steps to the right of the origin become significantly more intricate and an explicit expression cannot be 
found, The asymptotic behaviour, however, is readily evaluated: it is an exponential decay to zero of the 
concentration of the minority species (in this case, the left-going particles) and a corresponding exponential 
saturation of the majority species concentration to an asymptotic non-zero value of $p-q$. This exponential
approach is further corrected by a power-law factor: 
\begin{equation}
S(t)-S_\infty\simeq(pq)^{1/4}(\rho vt)^{-3/2}\dfrac{\exp(-2r\rho vt)}{r\sqrt{8\pi}}
\label{eq:2.3}
\end{equation}
where $r=1-2\sqrt{pq}$ is a quantity that measures the strength of the bias. 

For the case $p=q=1/2$, one can physically argue the decay law at long time as follows: $S(t)$ is mostly determined by 
the difference between the positive and negative velocity particles in an interval of time $2vt$. As the 
particles are initially distributed over the lattice independently with density $\rho$, the difference 
between the positive and negative velocity particles is of the order of $(2\rho vt)^{-1/2}$ for the population of order $2\rho vt$. 
 Hence the concentration of the particles at time $t$ (for $t\rightarrow \infty$) is given by 
 \begin{equation}
  c(t)=\rho S(t) \simeq \sqrt{\dfrac{c(0)}{\pi vt}},
  \label{concsyn}
 \end{equation}
where $c(0)= \rho$ is the initial concentration of the particles. 
 
 We can expand the argument given in the previous paragraph to give a simple derivation for the concentration 
 decay using scaling analysis \cite{baldif}. If the size of the one dimensional lattice is $L$, the initial number of particles 
 are $c(0)L$. The 
 difference between the number of left-moving and right-moving particles is of the order of 
 \begin{equation}
\Delta N = |N_+ - N_- | \sim \sqrt{N}.
\end{equation}
At long time ($t \rightarrow \infty$) all the particles which belong to the minority-velocity species are 
annihilated and hence $c(L,t\rightarrow \infty) \sim \Delta N/L \sim (c(0)/L)^{1/2}$. We can assume a scaling form 
for the concentration 
\begin{equation}
c(L,t) \sim (c(0)/L)^{1/2}g(z), 
\end{equation}
with $z=L/vt$. Following the above argument, we have
\begin{equation}
g(z) \rightarrow \mbox{\rm constant}\qquad(t\to\infty,z\to0).
\end{equation}
On the other hand in the short time limit, that is for $z \rightarrow \infty$,
the concentration cannot depend on the size of the lattice, which implies 
\begin{equation}
g(z) \propto z^{1/2}
\end{equation}
Thus we
find  
\begin{equation}
c(t) \sim \sqrt{\dfrac{c(0)}{ vt}}
  \label{cfnlsyn}
  \end{equation}
an expression similar to that of the solution given by equation (\ref{concsyn}).
\subsection{The case of three velocities}
\label{susec:3v}
Matters become rather more intricate in the case of three velocities. If we have particles moving
with velocities $v_1$, $v_ 2$, and $v_3$ we may suppose, using the Galilean invariance of the process,
that one of the velocities is equal to zero, say the intermediate velocity. To limit the number
of parameters, let us assume that the two extreme velocities are opposite, that is, we have
velocities $\pm v$ and $0$. Let the probability distribution be
 \begin{widetext}
\begin{equation}
D(w)=p_-\delta(w+v)+p_0\delta(w)+p_+\delta(w-v)\qquad(p_-+p_0+p_+=1).
\label{eq:2:4}
\end{equation}
 \end{widetext}
This system had been studied numerically in \cite{ball3v}. Qualitatively, the behaviour is the following: whenever the concentrations
of one type of particles clearly dominates all others, only those particles that have higher concentrations will eventually survive. The phase diagram is therefore
divided in three phases, in which either the $+$ particles, the $-$ particles or the immobile particles survive. Each phase has a 
boundary with the other two, and the three have only one point in common, which is numerically found to be very close
to $p_\pm=3/8$ and $p_0=1/4$. 

We now describe the behaviour of each species in different regions of the phase diagram. 
Within the regions in which only $\pm$ or immobile particles survive, the approach to equilibrium
is exponential. In the boundary separating the $+$ from the $-$ region, we have
\begin{equation}
c_\pm(t)\simeq t^{-1/2}\qquad c_0(t)\simeq t^{-1}. 
\label{eq:2.5}
\end{equation}
In other words, all particles disappear, but the immobile particles do so much faster than the moving
particles. If this is assumed, it is clear that the decay of the $\pm$ particles will asymptotically
not be affected by
the immobile ones, so that the problem is reduced to the case of two velocities treated in the previous section.
On the boundaries separating the $+$ or the $-$ phase from the immobile phase, the behaviour is unclear. 
Finally, at the tricritical point where all regions meet, namely $p_0=1/4$ and $p_\pm=3/8$, all particles again disappear,
but all with the same exponent, which is found to be $2/3$ with high numerical accuracy. 

These same results were later confirmed analytically \cite{ball3v1, ball3v2}. The essential argument described in \cite{ball3v1}
involves the factorization of the distribution of velocities to the left and to the right of any given particle. This implies that
the nearest neighbour distance distributions obey a {\em closed\/} hierarchy of equations. Note that this is similar, but
by no means identical, to the usual hierarchies involving {\em correlation functions} \cite{hierarchy}. The distribution of nearest-neighbour distances
involve correlation functions of arbitrarily high order. The quantities used in this approach are the nearest-neighbour distributions, and
these factor for reasons related to those we have already discussed.

The approach is not straightforward, however. In principle, it appears that any ballistic aggregation system can be in principle 
analyzed in this manner. However, even the detailed analysis of the three-velocity ballistic annihilation problem studied numerically
in \cite{ball3v} is an amazingly involved {\em tour de force}. An altogether different problem is the case of ballistic aggregation
in the case of {\em continuous\/} velocity distributions. This was treated within a Boltzmann equation approximation
in \cite{ballcontinuous}.  The results fit closely the numerics.

\section{Ballistic Annihilation with superimposed diffusion}
\label{sec:ball:async}
\subsection{Description of the model}
\label{subsec:intro}
In this section, we review the problem of ballistic annihilation with superimposed diffusion for a
the two-velocity case, in which particles can have only the velocities, $\pm v$
Initially the fraction of positive velocity particles are $p=1/2 + \varepsilon$ and the negative velocity particles 
are $q=1/2-\varepsilon$. The particles are randomly distributed over the one dimensional
lattice as usual. Most importantly, additionally to the ballistic particle motion we have a diffusive motion 
for the particle. This may, for instance, arise 
from an {\em asynchronous\/} update for the Monte Carlo Simulation. At  each  time  step  one  particle  is  chosen  
at  random  and moved in the direction of the corresponding velocity 
of that particle. After $N$ such update one Monte Carlo time step is over, if the total number of particles over 
the lattice is $N$. We consider the one species annihilation, that is whenever two particles meet they annihilate 
($A+A \rightarrow \emptyset$) \cite{baldif, shf}. 

The crucial difference between this model and the purely ballistic model is the possible annihilation between
particles having the same velocity. In other words, the purely ballistic model involves two particle species 
(left-going and right-going) which cannot react. In the model with diffusion, however, the particles of the same species
have another mechanism, namely diffusion, that allows them to annihilate. Since the two mechanisms scale differently, however,
the resulting model is in a different universality class from either the ballistic or the purely diffusive model. A trivial example of the difference is 
given by the case in which $p$ is significantly larger than $q$: in a time of order one, the purely ballistic model reaches a limiting
concentration, whereas the model involving diffusion decays as $t^{-1/2}$.

For $\varepsilon=0$,  that is, for $p=q$, at the beginning, the  numbers  of  positive  and  negative  velocity 
particles are equal on an average. When $\varepsilon \neq 0$ ($-1/2 \leq \varepsilon \leq 1/2$) introduces 
inequality in the initial numbers of positive and negative velocity particles, and $\varepsilon = \pm 1/2$ means there is 
only one kind of particle. 

 Unlike in the previous section, we study this problem in discrete time and at each time step we choose a particle at 
 random and move it one step to the right (or left) if the velocity of the particle is positive 
(or negative). In the asynchronous update, the random choice of particle leads to an effective diffusive motion of the 
particles with respect to their neighbouring particles that have the same speed. This asynchronous
update rule introduces diffusion so that, as remarked above, two particles with the same velocity can annihilate. 

\begin{figure}[ht]
 \includegraphics[width=5.5cm,angle=270]{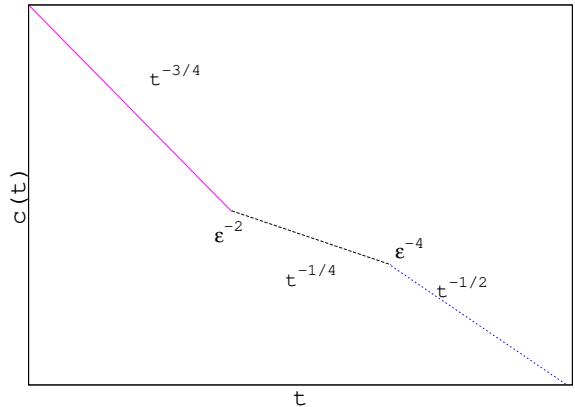} 
  \caption{A schematic picture with two crossovers for the decay of the concentration $c(t)$ for the total number of particles with time $t$.}
  \label{schematic_asyn}
\end{figure}

Let us here summarize the effects of this apparently minor change in the update rule.
These are non-trivial and profound.
 For $\varepsilon > 0$ ($\varepsilon \ll  1$), there are
three different regimes [Figure. \ref{schematic_asyn}].
 In the first stage we can neglect the
concentration difference between left- and right-going particles. The common concentration $c(t)$ 
then decays, as we will show, as 
$t^{-3/4}$. This regime ends when the minority ( that is left-going) particles begin to have as significantly different concentration
fro the majority. This is followed in short order by their complete disappearance. 
This happens at a time $t_1(\varepsilon)$ of the order of $\varepsilon^{-2}$ [Fig \ref{schematic_asyn}].  
At this point, starts the second stage which is characterized by the decay of concentration as $t^{-1/4}$
[Fig \ref{schematic_asyn}]. This is due to the fact that the distribution of the positions of the
surviving right
going particles over the one dimensional lattice are far from being uniform, 
leading to an anomalous decay \cite{alemany}. Finally, a third stage set in, when the distribution of the particles
over the lattice becomes uniform due to the presence of diffusion in the system, which happens at a 
time $t_2(\varepsilon)$ that scales as $\varepsilon^{-4}$ [Fig \ref{schematic_asyn}]. In this third stage, 
the system contains only the right-moving particles moving ballistically, with superimposed  diffusion.
Due to Galilean invariance, this is equivalent to pure diffusive dynamics. Hence the
concentration decay as $(Dt)^{-1/2}$ asymptotically. 

Even for $\varepsilon=0$, there will be an inequality in the right and left going
particle due to the random distribution of the particles at the beginning. For 
$\varepsilon =0 $,the concentration of an excess number of particles, $c_{ex}(0) 
\simeq 1/\sqrt{N}$ (up to an $O(1)$ factor) where $N$ is the initial number of
  particles. Hence we can see both the first and second stage there, but the
  crossover time diverges with the system size. In the next subsection we will
  discuss the initial $t^{-3/4}$ decay using dimensional analysis mostly for 
  $\varepsilon = 0$. 
  
\subsection{Dimensional analysis for $\varepsilon=0$ : Infinite system size}
\label{subsec:infinite}
Due to the convection, after some time, particles organize themselves into domains 
of right and left-moving particles.
However 
because of asynchronous update rule, inside each domain of same-velocity particles, 
diffusive annihilation takes place. One could therefore with some plausibility 
argue that the diffusive annihilation 
mechanism leads to an effective time-dependent initial concentration
$c_0(t) \sim (Dt)^{-1/2}$, which plays the role of $c(0)$ in Equation (\ref{cfnlsyn}),  
leading to the $t^{-3/4}$ decay of the concentration with time. 

However in this subsection 
we present a more thorough approach to this system. Specifically, we shall use 
dimensional analysis, or scaling, to derive the $t^{-3/4}$ decay law, for $\varepsilon =0$.
We follow the argument presented in
\cite{shf}. We define two dimensionless parameters:
 \begin{subequations}
 \begin{eqnarray}
x=\dfrac{v}{Dc(0)}
\label{dimlessa}\\
\tau=\dfrac{v^2t}{D}
\label{dimlessb}
\end{eqnarray}
\label{eq:dimless}
 \end{subequations}
where $c(0)$ is the initial concentration, $D$ is the diffusion constant and $v$ is the velocity of the 
particles. The meaning of these quantities is as follows: $x$ is the ratio of two times: the time
needed to cross the initial inter-particle separation ballistically, and that needed to cross the same distance 
diffusively. O the other hand, 
 $\tau$ is a dimensionless time (Equation \ref{dimlessb}), given by the ratio of $t$
to the time scale $\tau_{v}$ above which ballistic motion dominates diffusion. Hence at times
corresponding to $\tau\ll1$, diffusion
dominates, whereas for $\tau\gg1$, drift does. 
 
 Now the concentration of the particles as a function of $x$ and $\tau$, at time $t$ can be written as :
 \begin{equation}
  c(t) \simeq c(0)\Phi (x, \tau )
 \end{equation}

If $x \ll 1$, $c(t)$ cannot depend on $v$ at small times, since then $\tau\ll1$ and everything 
is dominated by diffusion. That implies that $c(t) \sim (Dt)^{-1/2}$ for small $t$. Hence using Equation (\ref{dimlessa})
 and (\ref{dimlessb}) we get 
\begin{equation}
\Phi(x, \tau)=\dfrac{x}{\sqrt\tau}
\label{weakbias}
 \end{equation}
This should be valid until $\tau \sim 1$, as this is the crossover time between drift and diffusion. 

Now consider the case $x \gg 1$. In that case, since time is discrete, $\tau\gg1$ for all times. It may therefore appear
that the dynamics is purely ballistic since $\tau\gg1$; $c(t)$ is thus independent
of $D$ in this regime,  and it is  given by $\sqrt{c(0)/vt}$, as discussed in the previous section. In this case, again
using Equation  (\ref{dimlessa}) and (\ref{dimlessb}), we get 
\begin{equation}
\Phi(x, \tau)=\sqrt{\dfrac{x}{\tau}}.
\label{strongbias}
\end{equation}
The structure resulting from this dynamics is therefore that of the ballistic aggregation without diffusion
studied in the previous Section. I this model, however, there form large domains of particles moving at the same velocity, 
which do not react in the pure ballistic model.

Now, due to the presence of the diffusion, the particles forming such domains can annihilate via diffusion. Since 
the particles in a domain have never interacted, we may roughly assume their separation to be given, in terms
of the initial concentration, by $1/c(0)$. Particles having the same speed therefore 
react on a time scale $1/(Dc(0)^2)$, which in terms of the dimensionless quantities, is $\tau\sim x^2$.
Hence Equation (\ref{strongbias}) will be valid until time $\tau\sim x^2$. The validity of (\ref{strongbias})
is therefore limited to finite, though large, values of $\tau$. 

As explained above, the previous equations (\ref{weakbias}) and (\ref{strongbias}) only hold for finite values
of $\tau$.
Now let us describe the large $\tau$ behaviour for the full dynamics. In that case we can write 
\begin{equation}
\Phi(x, \tau)\simeq x^\alpha\tau^\beta
\label{asympt}
\end{equation}
When $\tau\sim 1$ and $x\ll1$, we can use both equation (\ref{weakbias}) and (\ref{asympt}), which leads to
\[
x\sim x^\alpha
\]
That gives $\alpha=1$. Similarly, if $x\gg1$ and $\tau\sim x^2$, we can use the equation (\ref{strongbias}) 
as well as (\ref{asympt}), which lead us to 
\[
\sqrt{\dfrac{x}{\tau}}\sim x^{-1/2}\sim x^{\alpha+2\beta}=x^{1+2\beta},
\]
and that give us 
\[
 \beta = -\dfrac{3}{4}.
\]
Putting the value of $\alpha$ and $\beta$ in Equation \ref{asympt} we get 
\[
 \Phi(x, \tau)\simeq x\tau^{-3/4}.
\]
From this we finally get that for large time ($t\rightarrow \infty$) the concentration of the particles 
$c(t)$ will have the following form : 
\begin{equation}
 c(t)\simeq v^{-1/2}D^{-1/4}t^{-3/4}.
\label{powerz1}
\end{equation}

Now we will see an alternative way to determine the decay law using dimensional analysis in a little 
different way. As we have already discussed the system can be completely characterized by the initial 
concentration $c(0)$, velocity $v$ and diffusion constant $D$. From these parameters, one can get different
combinations with the dimensions of concentrations, which are $c(0)$, $1/vt$ and $1/\sqrt{Dt}$. From this one 
can anticipate that these three concentration scales could be written multiplicatively to express the 
time-dependent concentration in a conventional scaling form. Hence we can write the time-dependent 
concentration as follows : 
\begin{equation}
 c(t) \sim c(0)^{\rho} \left(\dfrac{1}{vt} \right)^{\sigma} \left(\dfrac{1}{\sqrt{Dt}} \right)^{1-\rho-\sigma}
\label{dimless}
\end{equation}
The exponents $\rho$ and $\sigma$ can be obtained by the following argument: first, let $t < \tau_v \simeq D/v^2$.
Here $\tau_v$ is the characteristic time below which the diffusion dominates over drift. The 
particle then undergoes (weakly) biased diffusion, which leads to 
\begin{equation}
c(t) \sim (Dt)^{-1/2}.
\end{equation}
independently
of $c(0)$. At $t=\tau_v$, putting the expression 
of $c(t)$ and $\tau_v$ in equation (\ref{dimless}), we obtain $\rho =0$. Secondly, consider $\tau_v<t < \tau_D \simeq 1/(Dc(0)^2$, 
where $\tau_D$ is the time needed for two neighbouring same velocity particles to meet by diffusion. In this case 
the dynamics is mostly ballistic and hence 
\begin{equation}
c(t) \sim \sqrt{c(0)/(vt)}.
\end{equation}
 Again at $t=\tau_D$, putting the expression 
of $c(t)$ and $\tau_D$ in equation (\ref{dimless}), we obtain $\sigma =1/2$. Using the value of $\rho$ and $\sigma$ 
in equation (\ref{dimless}), one can get back the same expression for $c(t)$ at large time which is given by 
equation (\ref{powerz1}).

 Note that so far we have not at all taken into account the possibility of unequal concentrations of left-going and right-going particles.
Similarly, we have not considered the effect of a finite system size. 
We now turn to these issues.
\subsection{The case of finite system size : Second stage of dynamics}
\label{subsecfrac}

In an infinite system, it is readily seen that the arguments of Subsection \ref{subsec:infinite} hold
for all times. For finite systems, another regime appears, which we now describe.

When all the minority particles have been annihilated, the {\em only\/} reaction mechanism remaining 
is diffusive annihilation. One might thus be tempted to conclude that, for finite systems, we merely
have ab additional final stage in which ordinary diffusive annihilation occurs, with a decay of $1/\sqrt{Dt}$. 
As we show, matters are a bit less straightforward.

At
the beginning of this stage, that is, when all minority particles have disappeared, 
the distribution of all the surviving particles (which are a fraction of the 
majority particles) over the one dimensional lattice is far from being uniform. This leads 
to the next decay law. In this subsection we describe by simple arguments, the
spatial distribution of the particles over the one dimensional lattice, and argue that 
it leads to $t^{-1/4}$ decay.
We  follow the arguments given in \cite{shf}, to which the reader is referred for details. 

Let us start with the behaviour of the surviving particles in ballistic annihilation with
synchronous updating. Once the initial condition is set at random, each particle has a unique annihilation partner. Otherwise the particle survives forever. Initially the distance between the particle (positive
velocity) situated at the $k^{th}$ lattice cite and the unique annihilation partner of that particle be 
$\pi_+(k)$. $k$ runs from $0$ to $L$, where $L$ is the length of the one dimensional lattice. Once the
initial configuration is fixed each particle survives until it encounters its reaction partner. Hence the
collision time is
\begin{equation}
\tau_+(k)=\dfrac{\pi_+(k)}{2}
\label{eq:tcolision}
\end{equation}
Here we have considered the positive velocity particles. The entire description is also valid for
 the negative velocity particles. In that case we would say that the initial distance of the unique
 annihilation partner be $\pi_-(k)$.
 
 We can now try to determine the structure of the set $\Sigma_t$, which is defined as
\begin{equation}
\Sigma_t=\left\{
k:\tau_+(k)>t
\right\}
\label{eq:3}
\end{equation}
Let us consider that $k$ is the distance between two positive velocity particles. Without loss of 
generality we may assume that we have two particles, one at $0$ and the other one at $k>0$. 

If $k>t$, then the two intervals $[0,t]$ and $[k,k+t]$ are disjoint. The probability that both sites 
belong to $\Sigma_t$ is simply the product of either site belonging to $\Sigma_t$ and hence no dependence on
$k$ exists. This is independent of the fact whether we consider $\varepsilon=0$ for the initial configuration or not.

On the other hand, for $k<t$ the situation is little different. Let us assume a
virtual random walk that takes a step to the right whenever there is positive
velocity particle in the initial configuration and takes a step to the left if the
velocity of the particle is negative. For $\varepsilon =0$, the walk is symmetric.
Hence for $\varepsilon=0$, the probability of having two particles separated by a 
distance $k<t$ and both surviving a time $t$ is equivalent to the the probability
that a symmetric random walk, which starts at the origin and takes a step to the 
right does not return to the origin before time $k$. This probability
scales as $k^{-1/2}$ for $k\gg1$, if the walk is symmetric \cite{Weiss}. This 
description is compatible with the idea that the set $\Sigma_t$ forms a fractal
set---below the cutoff value $t$ and with fractal dimension $1/2$.

For synchronous update we see that the set $\Sigma_t$ is a fractal with a lower 
cutoff at length $1$ (lattice spacing) and an upper cutoff $t$ with fractal 
dimension $1/2$. In case of asynchronous update, annihilation of the same velocity
particles due to the the presence of diffusion, eliminates all (well, actually, most of) 
the particles with distance less than $\sqrt t$. Hence the corresponding set of 
surviving particles becomes a fractal set of dimension $1/2$ with lower cutoff
$\sqrt t$ and upper cutoff $t$. 

For $\varepsilon > 0$, the argument expanded so far is still valid. The main
difference is that now the virtual random walk is biased. That virtual walker now 
takes a step to the right with probability $1/2+\varepsilon$, whereas a step to the left has probability $1/2-\varepsilon$. For $k<t$, the probability that two 
particles, one at $0$ and another at $k$ both belong to $\Sigma_t$, is still given
by the probability (say $p_\varepsilon(k)$), that a random walk (biased now), which
starts at the origin and takes a step to the right, does not return to the origin
before time $k$. 

 The probability $p_\varepsilon(k)$  {\em saturates\/} to a positive value 
 $p_\varepsilon(\infty)$ as $k \rightarrow \infty$. This saturation happens when 
 $k\sim k_c(\varepsilon) \sim\varepsilon^{-2}$ as $\varepsilon\to0$ \cite{Weiss,
 Hughes} and when $\varepsilon \rightarrow 0$ we have $p_\infty(\varepsilon)\simeq
 \varepsilon$. The set $\Sigma_t$ is therefore a fractal set with a cutoff which is 
 either at $t$ or at $\varepsilon^{-2}$, whichever is smaller.

From the arguments given above we see that the particles surviving at time $t_1$
(Fig. \ref{schematic_asyn}), at which only one species (either $+ve$ or $-ve$ 
 velocity particles) survives, lie on a fractal of dimension $d_f=1/2$. This is 
 also true for both $\varepsilon=0$ and $\varepsilon > 0$ in the initial condition.
 
 The fact that $\Sigma_{t_1}$ is a fractal further implies that the second stage 
 of the dynamics is purely diffusive (as only one species survives), starting from an initial 
 condition where the particles are distributed on a fractal of dimension $d_f$ ($0<d_f<1$). 
 The probability distribution function for this initial
 inter-particle distances for two nearest particles has,  for $x\gg1$, a behavior
 similar to that of the distribution
\begin{equation}
 P(x)=\dfrac{1}{\zeta(\lambda)}\sum_{l=1}^{\infty} l^{-\lambda}\delta(x-l)
\label{indist}
\end{equation}
where $\lambda = d_f +1$. For the purposes of the asymptotic estimations we shall be dealing
with, $P(x)$ and the exact probability distribution can be used interchangeably. 
Considering this discrete distribution given by equation
(\ref{indist}), and following  the  formalism  developed in \cite{alemany}, one can
obtain the number of particles $n(t)$ at time $t$ (normalized by initial number of
particles $n(0)$) which is given by 
\begin{eqnarray}
 n(t)=-\dfrac{\Gamma(1-\lambda)}{2\zeta(\lambda)\Gamma(\dfrac{3-\lambda}{2})} \tau^{-\dfrac{\lambda -1}{2}} + 
 \dfrac{\Gamma(1-\lambda)}{4(1-\lambda) [\zeta(\lambda)]^2}~\tau^{-(\lambda-1)}
 \nonumber \\ 
+ \dfrac{\zeta(\lambda -1)}{2\zeta(\lambda)\sqrt{\pi}}~\tau^{-1/2} + O(\tau^{-\lambda/2})~~~~~~
\label{nwfrct}
\end{eqnarray}
where $\tau=D_{eff}~t$. See Appendix in \cite{shf} for details. In our case the 
fractal dimension $d_f=1/2$, and hence $\lambda=3/2$. The leading term of the 
decay law is $t^{-1/4}$ and the coefficient of this leading term is positive, as 
$\Gamma(1-\lambda)$ is {\em negative\/} for all $\lambda >1 $. Putting the value of 
$\lambda$ in equation (\ref{nwfrct}) we get
\begin{eqnarray}
n(t)=\dfrac{\sqrt\pi}{\zeta(3/2)\Gamma(3/4)}\tau^{-1/4}+ \nonumber \\
\left(\dfrac{\sqrt{\pi}}{\zeta(3/2)^2}+\dfrac{\zeta(1/2)}{2\sqrt\pi\,\zeta(3/2)}
\right)\tau^{-1/2} 
+O(\tau^{-3/4})\nonumber\\
\approx
0.5537\,\tau^{-1/4} + 0.102\,\tau^{-1/2}+O(\tau^{-3/4})
\label{nw05frc}
\end{eqnarray}

Now let us progress to determine the crossover times $t_1(\varepsilon)$ and 
$t_1(\varepsilon)$. $t_1(\varepsilon)$ is the time at which the dynamics have a 
crossover from $t^{-3/4}$ initial behaviour, to the long-time $t^{-1/4}$ behaviour.
Hence at $t_1(\varepsilon)$ we can write
\begin{equation}
 at_1(\varepsilon)^{-3/4}=b(\varepsilon)t_1(\varepsilon)^{-1/4}
 \label{scrvr}
\end{equation}
where the coefficient $a$ does not depend on $\varepsilon$, as the decay at the
beginning does not depend on the initial concentration. However $b(\varepsilon)\sim \varepsilon $, as $p_\infty(\varepsilon)\simeq\varepsilon$ for $\varepsilon
\rightarrow 0$.  Simplifying equation (\ref{scrvr}), we obtain
\begin{equation}
 t_1(\varepsilon) \sim \varepsilon^{-2} 
 \label{t1cr}
\end{equation}
For $\varepsilon \rightarrow 0$, this crossover time $t_c$ will diverge and hence
will scale with system size $L$ for finite $L$. We will still be able to see the
$t^{-1/4}$ decay due to the concentration of an excess number of particles, 
$c_{ex}(0) \simeq 1/\sqrt{N}$ for setting the initial configuration at random.

Though it is possible to see the $t^{-1/4}$ decay for $\varepsilon=0$ it is 
difficult  as there remain very few particles at this late stage.  However 
$\varepsilon \neq 0$, it is respectively easier to detect the $t^{-1/4}$ (which is
the leading term) decay of concentration on the fractal, as the remaining particles
are more in this situation. As the dynamics is entirely diffusive at this stage,
the fractal structure will eventually die out to uniform distribution. So there 
will be another crossover after which the concentration will decay as $t^{-1/2}$ 
at very long time. If we consider the crossover time for the second crossover as
$t_2(\varepsilon)$ we can write
 \[
b(\varepsilon)t_2(\varepsilon)^{-1/4}=c\,t_2(\varepsilon)^{-1/2}
 \]
where the coefficient $c$ is not a function of $\varepsilon$ as it does not depend
on the initial concentration. Simplifying the above expression we get 
\begin{equation}
 t_2(\varepsilon) \sim \varepsilon^{-4}
 \label{t2cr}
\end{equation}
It is indeed rather challenging to measure this second crossover time $t_2(\varepsilon)$,
as the number of remaining particles is very low and the time very large.

In Subsection \ref{subsec:scaling} we will show some of the numerical evidence supporting
the theoretical claims described in this subsection and in Subsection \ref{subsec:infinite}. 

\subsection{Numerical Evidences and scaling}
\label{subsec:scaling}
If initially there are $N$ number of particles, then at time $t_1(\varepsilon)$
there remain $N\varepsilon^{3/2}$ particles and at $t_2(\varepsilon)$ the number of particles remain only $N\varepsilon^{5/2}$. This makes some numerically challenging
features of the model: if we wish to have a clean separation of time scales, we
need at least $\varepsilon\sim10^{-2}$. In a different language, if we wish to
have more or less hundred particles at time $t_2(\varepsilon)$ (for being able to
 observe the final exponent reliably), we need to start with $N\sim10^7$, which is 
 unrealistic. Hence to drag out the features of the system explained in the 
 previous subsection one have to rely on various simulations with different values
 of the parameters and the corresponding scaling laws. 
 
\begin{figure}[ht]
\includegraphics[width=5.7cm,angle=270]{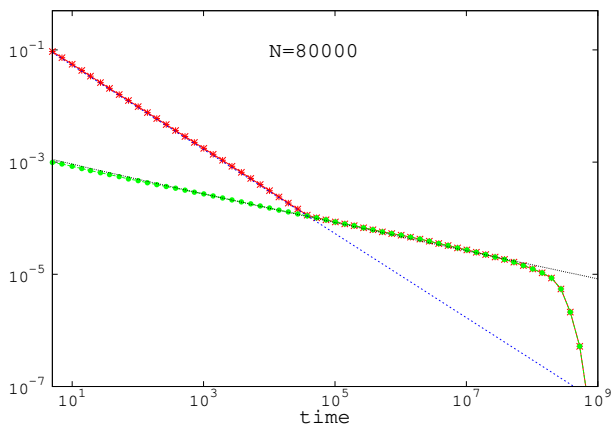}
\caption{ The concentration $c(t)$ (red points) for the total number of particles
  decay as $t^{-3/4}$, (the slope is given by blue dotted line) before the
  crossover and following the equation (\ref{nw05frc}) after the crossover. Decay for the concentration for the excess number of particles $c_{ex}(t)$ is plotted
  in green points. $n(t)/\sqrt{N}$, with $\lambda=1.5$ is plotted as the
  theoretical curve (black dotted line), where the expression of $n(t)$ is
  given by equation (\ref{nwfrct}). The decay of the concentrations
  is plotted starting with initially $N=80000$ particles.}
\label{nw}
\end{figure}

 First we see the decay of concentration for $\varepsilon=0$ which shows the 
 $t^{-3/4}$ decay law at the beginning and then a crossover to $t^{-1/4}$ decay
 (Fig. \ref{nw}). 
As we know, initial concentration for the excess particles $c_{ex}(0)=1/\sqrt{N}$.
If the excess particles, which decay due to diffusive annihilation lie on a fractal
of dimension $1/2$ from the very beginning of the dynamics, then they will decay
following the equation (\ref{nw05frc}). The plot of Equ. \ref{nw05frc} with the 
numerical data for $c_{ex}(t)$ (Fig. \ref{nw}) shows perfect agreement, supporting this assumption.

\begin{figure}[ht]
\includegraphics[width=8.6cm,angle=0]{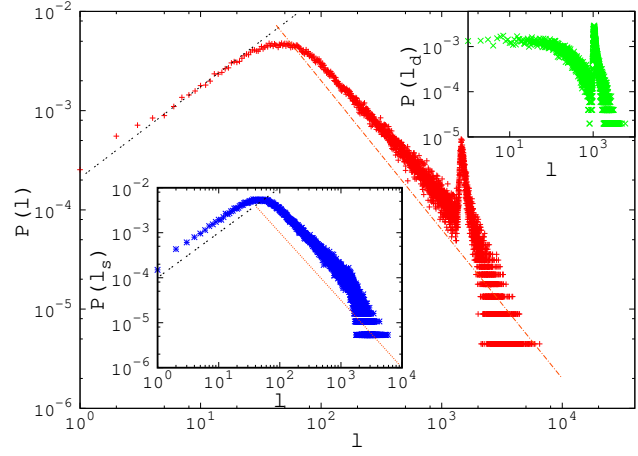}
\caption{Main plot shows the general distribution $P(l)$ for the inter-particle
distances between two neighbouring particles. Inset at the bottom shows the
distribution $P_s(l)$ and the top right inset shows the distribution $P_d(l)$. Saffron line have slope $l^{-3/2}$ for both the main plot and at the bottom inset. Black dashed line shows the linear increase. }
\label{distnn}
\end{figure}

Next we will show the distribution of inter-particle distances between two
neighbouring particles at late times, when the particles decay as $t^{-1/4}$. 
$P_s(l)$ and $P_d(l)$ are the distribution functions of inter-particle distances
between two neighbouring particles with same and different velocities [Fig. \ref{distnn}]. 
$P_s(l) \sim l^{-3/2}$ for large $l$ [Inset at the bottom of Fig.\ref{distnn}], 
which is evidence that inside a domain of same velocity particles, the particles 
are distributed over a fractal of dimension $1/2$. For small $l$, $P_s(l) \sim l$
due to the diffusion-limited annihilation \cite{DBA}. The distribution $P_d(l)$ 
is almost flat at the beginning and then has an exponential decay [Top right
inset of Fig. \ref{distnn}]. 
At the part of the exponential decay the value of $P_d(l)$ suddenly increases as
two domains or fractals of same velocity particles are moving apart from each 
other making the probability of having some large value of $l$ very high.
$P(l)$ is the general distribution function for inter-particle distances
between two neighbouring particles, independent of their velocity [Main plot of \ref{distnn}].

Next we will show the numerics for $0<\varepsilon<1/2$. Though for 
$\varepsilon=0$, there exist three different dynamical regimes, due to the
finiteness of the system, the regimes are not that well separated and hence the
crossover times are not very clean. We will mostly discuss the scaling behaviours
 involving $\varepsilon$ and $t$, which hold for $|\varepsilon| \ll 1$, for two
 crossovers in this article. For details of the numerics one should consult 
 \cite{shf}. 

\begin{figure}[ht]
\includegraphics[width=8.6cm,angle=0]{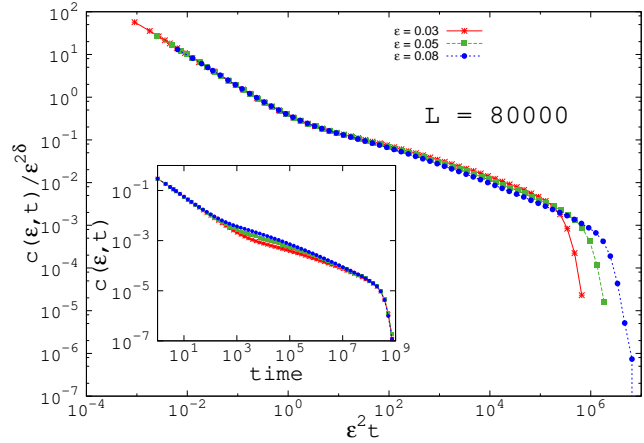}
\caption{The main plot shows the collapse of scaled data of concentration of
  particles with $\delta=3/4$ for different $\varepsilon \neq 0$.
  Inset shows the raw data. }
\label{nw_1_colp}
\end{figure}

The scaling function for the first two dynamical regimes can be written as
\begin{equation}
 c(\varepsilon,t) \sim \varepsilon^{2\delta} f(\varepsilon^2t) ,\qquad\hbox{\rm for }\varepsilon \ll 1
 \label{fss1_epsi}
\end{equation}
where $f(x) \rightarrow x^{-\delta}$ with $\delta=3/4 $, for $x \ll 1$ and 
$f(x) \rightarrow x^{-1/4}$ when $x \gg1$.  The raw data as well as the scaled
data using $\delta=3/4$ are shown in Fig. \ref{nw_1_colp}. The collapse is good
for the first two dynamical regimes where the exponent values are $3/4$ and
  $1/4$ respectively.

The scaling law for the second and third dynamical regime (where $\varepsilon^4t$
is the relevant dimensionless quantity) can be written in the following form
\begin{equation}
 c(\varepsilon,t) \sim \varepsilon^{4\eta} g(\varepsilon^4t),\qquad\hbox{\rm for all }\varepsilon \ll 1
 \label{fss2_epsi}
\end{equation}
where $g(x) \rightarrow x^{-\eta}$ with $\eta=1/2 $, for $x
\rightarrow \infty$ and $g(x) \rightarrow x^{-1/4}$ when $x \ll1$.

\begin{figure}[ht]
\includegraphics[width=8.6cm,angle=0]{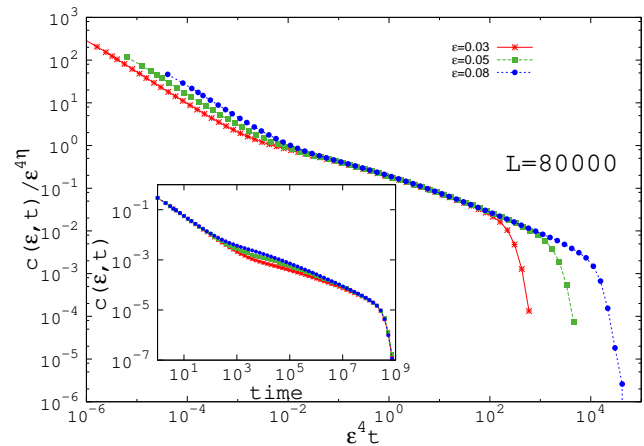}
\caption{(Collapsed plot of scaled data for the concentration of
  particles with $\eta=1/2 $ for different $\varepsilon \neq 0$.
  Inset shows the raw data.   The average number of
  particles is again considerably less than $one$ in this
  exponentially decaying region. }
\label{nw_2_colp}
\end{figure}
The quality of collapse of the scaled data is good for second and third dynamical
region where the exponent values are $1/4$ and $1/2$ respectively [Figure. \ref{nw_2_colp}]. Data collapse
can not be obtained for the exponentially decaying part, as the scaling theory
  applies only for the power law regions. 
  
  \section{Conclusions}
 \label{sec:conclusions} 
In this article we have reviewed the problem of ballistic annihilation with discrete velocities 
in one dimension. First we discussed the decay of the concentration of the 
particles under the synchronous update rule. After that we described the 
properties and characteristics of the dynamics when diffusion is superimposed, for instance
by making the update rule asynchronous.
There are several open questions on ballistic annihilation which could be addressed in future. One of the important 
issue that could be to studied is the consequence of initial particle distributions on the decay 
law for the ballistic annihilation with superimposed diffusion. If initially the particles are distributed over a 
fractal of fractal dimension $d\neq1$, then the concentration of the particles should still decay in a power law 
fashion, but with a different exponent depending on the value of $d$, the dimension of the system. 
Following the argument presented in the section \ref{subsecfrac},
the concentration decay should again go through a crossover in time, before it finally start decaying with exponent $1/2$. 
Checking this hypothesis could be a part of a future project. 

The problems of two species ballistic annihilation \cite{frnRchd} with superimposed diffusion in one and 
higher dimension are also open for the future study. On the other hand, the annihilation problem for the persistent 
random walkers \cite{persrw} is also very interesting that could be studied in future. 
One can view the problem of ballistic annihilation as a limiting case for $A+A \rightarrow \emptyset$, where $A$ walkers are the persistent walkers. 
This problem can be studied under both the synchronous and asynchronous update rule in future. The generalization of the above results
to three or more velocities is also a challenging problem.  

\section{Acknowledgement}

Financial support from the project of 
CONACyT Ciencia de Frontera 2019, Number \textbf{10872}, is gratefully acknowledged. FL acknowledges the financial support from the project of 
CONACyT Ciencias Basicas, Number \textbf{254515}. 

\textit{Author Contribution Statement:} Both the authors contributed equally to the paper.


\begin{thebibliography}{55}
\bibitem{review} For a review of diffusion-controlled annihilation, see 
S.~Redner, \textit{Nonequilibrium Statistical Mechanics in One Dimension} ed V.~Privman, Cambridge: Cambridge University Press (1996)

 \bibitem{book} P.L. Krapivsky, S. Redner and E. Ben-Naim, \textit{ A Kinetic View of Statistical Physics}; 
Cambridge: Cambridge University Press, (2010)

\bibitem{spouge} J. L. Spouge, ``Exact solutions for a diffusion-reaction process in one dimension'' Phys. Rev. Lett. \textbf{60} 871 (1988)

\bibitem{Francois}F. Leyvraz and N. Jan, ``Critical dynamics for one-dimensional models''
J. Phys. A: Math. Gen. \textbf{19} 603-605. (1986)

J. L. Spouge, ``Exact solutions for a diffusion-reaction process in one dimension: 11. Spatial distributions'', J. Phys. A: Math. Gen. \textbf{21}, 4183 (1988)

\bibitem{bnnni} S. Biswas and M. M. Saavedra Contreras, ``Zero-temperature ordering 
dynamics in a two-dimensional biaxial next-nearest-neighbor Ising model'' Phys. Rev. E \textbf{100}, 042129 (2019)

\bibitem{opn} I. Ispolatov, P.L .Krapivsky, S. Redner, ``War: The dynamics of vicious civilizations''
Phys. Rev. E \textbf{54}, 1274 (1996) 

S. Biswas and P. Sen, ``Model of binary opinion dynamics: Coarsening and effect of disorder'' 
Phys. Rev. E \textbf{80}, 027101 (2009) 

S. Biswas, P. Sen, and P.  Ray, ``Opinion dynamics model with domain size dependent dynamics: novel 
features and new universality class'' J. Phys.: Conf. Series \textbf{297}, 012003 (2011) 

\bibitem{balsyn} Y. Elskens and H. L. Frisch, ``Annihilation  kinetics in the one-dimensional ideal gas'', 
Phys. Rev. A \textbf{31}, 3812 (1985).

\bibitem{baldif} E Ben-Naim, S Redner and P L Krapivsky, 
``Two scales in asynchronous ballistic annihilation'',
J. Phys. A: Math. Gen. \textbf{29} L561 (1996)

\bibitem{feller}   William Feller, {\em An Introduction to Probability Theory and Its Applications}, John Wiley and Sons Ltd; 3rd edition (January 1, 1968).

\bibitem{ball3v} P. L. Krapivsky, S. Redner, and F. Leyvraz, ``Ballistic annihilation kinetics: The case of discrete velocity distributions''
Phys. Rev. E \textbf{51} 3977 (1995)

\bibitem{ball3v1} J. Piasecki, ``Ballistic annihilation in a one-dimensional fluid'' Phys. Rev. E \textbf{51} (6) 5535--5540 (1995)

\bibitem{ball3v2} M. Droz, P.-A. Rey, L. Frachebourg, and J. Piasecki, ``Ballistic-annihilation kinetics for 
a multivelocity one-dimensional ideal gas'' Phys. Rev. E \textbf{51} (6) 5541--5548 (1995)

\bibitem{hierarchy} S. Harris, \textit{An introduction to the theory of the Boltzmann equation}, Courier Corporation (2004)

\bibitem{ballcontinuous} E. ben-Naim, S. Redner y  F. Leyvraz,``Decay kinetics of ballistic annihilation'' 
Phys. Rev. Lett. \textbf{70} 1890 (1993)

\bibitem{shf} S Biswas, H Larralde and F Leyvraz, ``Ballistic annihilation with superimposed diffusion in one 
dimension''Phys. Rev. E \textbf{93}, 022136 (2016)

\bibitem{alemany} P. A. Alemany, ``Novel decay laws for the one-dimensional 
reaction-diffusion model as consequence of initial distributions'',  J. Phys. A: Math. Gen. 
\textbf{30} (1997) 3299

\bibitem{Weiss} G. H. Weiss, \textit{Aspects and applications of the random walk Random Materials and Processes}; ed H. E. Stanley and E. Guyon (1994)

\bibitem{Hughes} B. D. Hughes; \textit{Random Walks and Random Environments}, vol \textbf{1}; Oxford: Clarendon (1995)

\bibitem{DBA} C.R. Doering and D. Ben-Avraham, D.  ``Interparticle distribution functions and rate 
equations for diffusion--limited reactions''. Phys. Rev. A, \textbf{38} (6) 3035 (1988)

\bibitem{frnRchd} Yu Jiang and F. Leyvraz, ``Kinetics of two-species ballistic annihilation'' Phys. Rev. E \textbf{50}, 608 (1994)

M. J. E. Richardson, ``Exact solution of two-species ballistic annihilation with general pair-reaction probability''
J. Stat. Phys. \textbf{89}, 777 (1997)

\bibitem{persrw} J Masoliver, and G. H. Weiss, ``TelegrapherÕs equations with variable propagation speeds'' 
Phys. Rev. E \textbf{49}, 3852 (1994)

S. K. Foong and S. Kanno,  ``Properties of the telegrapher's random process with or without a trap''
Stochastic Processes Appl. \textbf{53}, 147 (1994).

G. H. Weiss,  \textit{Aspects and Applications of the Random Walk}, North-Holland, Amsterdam (1994)


\end{thebibliography}
\end{document}